\documentclass[final,1p,times]{elsarticle}

\usepackage{lineno}
\usepackage{amssymb}
\usepackage[figuresright]{rotating}

\journal{Nuclear Physics A}

\begin{document}
\begin{frontmatter}
\title{Open charm hadron production via hadronic decays at STAR}
\author{\small David Tlust\'y for the STAR collaboration}
\address[ujf]{Nuclear Physics Institute, Academy of Sciences Czech Republic, Na Truhl\'a\v{r}ce 39/64, 180 86 Praha 8, Czech Republic}
\address[cvut]{Czech Technical University in Prague, Faculty of Nuclear Sciences and Physical Engineering, B\v{r}ehov\'a 7, 11519, Prague 1, Czech Republic}
\begin{abstract}
In this article, we report on the STAR results of open charm hadron production at mid-rapidity in $p+p$ and Au+Au collisions at $\sqrt{s_{NN}}$ = 200 GeV and
$p+p$ collisions at  $\sqrt{s}$ = 500 GeV. The measurements cover transverse momentum range from 0.6 to 6 GeV/c for $p+p$ 200 GeV collisions, from 1 to 6 GeV/c for $p+p$ 500 GeV collisions
and from 0 to 6 GeV/c for Au+Au 200 GeV collisions. $D^0$ nuclear modification factor and elliptic flow in Au+Au collisions at $\sqrt{s_{NN}}$ = 200 GeV are presented. 
\end{abstract}


\end{frontmatter}

\linenumbers

\vspace{-5mm}
\section{Introduction}
\label{sec:intro}
\vspace{-2mm}
The heavy quark production at RHIC is dominated by initial gluon fusion at initial hard partonic collisions 
and can be described by perturbative QCD (pQCD) due to their large mass \cite{cprod}. The heavy constituent quark mass is almost exclusively generated through its coupling to the Higgs field in the electroweak sector, while masses of (u, d, s) quarks are dominated by spontaneous breaking of
chiral symmetry (CS) in QCD \cite{CharmMass}. This means that charm quarks remain heavy even if CS
is restored, as it likely is in a QGP. One expects therefore
that charm production total cross section
$\sigma^{NN}_{c\bar{c}}$ should scale as a function of number-of-binary-collisions $N_{\mathrm{bin}}$. 
In addition, if charm quarks participate in the collective expansion of the medium, there must have been enough interactions to easily thermalize light quarks. Hence, charm quark is an ideal probe to study early dynamics in high-energy nuclear collisions.      

\vspace{-3mm}
\section{Analysis Method and Datasets}
\label{sec:measurement}
\vspace{-2mm}

Invariant yield of charm quark production $\mathrm{Inv}Y$ is calculated as
\begin{equation}
\mathrm{Inv}Y\equiv\frac{\mathrm{d}^2N_{c\overline{c}}}{2\pi p_T\mathrm{d}p_T\mathrm{d}y}=\frac{1}{N_{\mathrm{trig}}}\frac{Y(p_T,y)}{2\pi p_T\Delta p_T \Delta y}\frac{f_\mathrm{trg}}{\mathrm{BR}\  f_\mathrm{frag.} \epsilon_\mathrm{rec}} 
\label{eq:InvariantYield} 
\end{equation}
where $N_\mathrm{trig}$ is the total number of triggered events used for the analysis. $Y(p_T,y)$ is the raw charm hadron signal in each $p_T$ bin within a rapidity window $\Delta y = 2$. BR is the hadronic decay branching ratio for the channel of interest. $\epsilon_\mathrm{rec}$ is the reconstruction efficiency including geometric acceptance, track selection efficiency, PID efficiency, and analysis cut efficiency. $f_\mathrm{frag.}$ represents the  the ratio of charm quarks hadronized to open charm mesons. And $f_\mathrm{trg}$ is  
the correction factor to account for the bias between the minimum-bias sample used in this analysis and the total NSD sample \cite{OpenCharm2011Paper}. $f_\mathrm{trg}$ is found to be unity in Au+Au, 0.65 in $p+p$ collisions at $\sqrt{s}=$ 200 GeV and 0.58 in $p+p$ collisions at $\sqrt{s}=$ 500 GeV.

\begin{figure}[!h]
\begin{center}
\vspace{-3mm}
\includegraphics[width=0.285\textwidth]{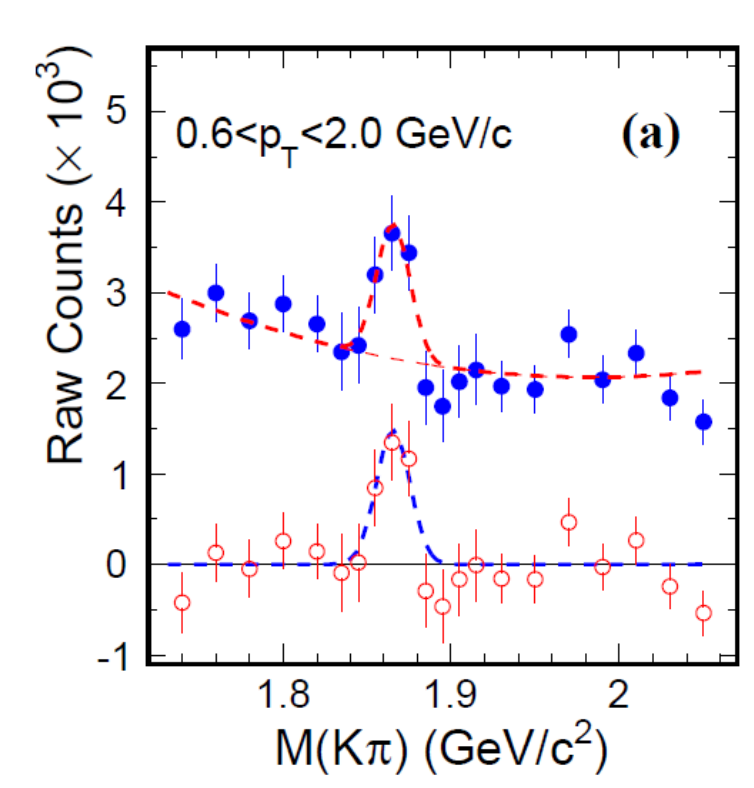} 
\includegraphics[width=0.255\textwidth]{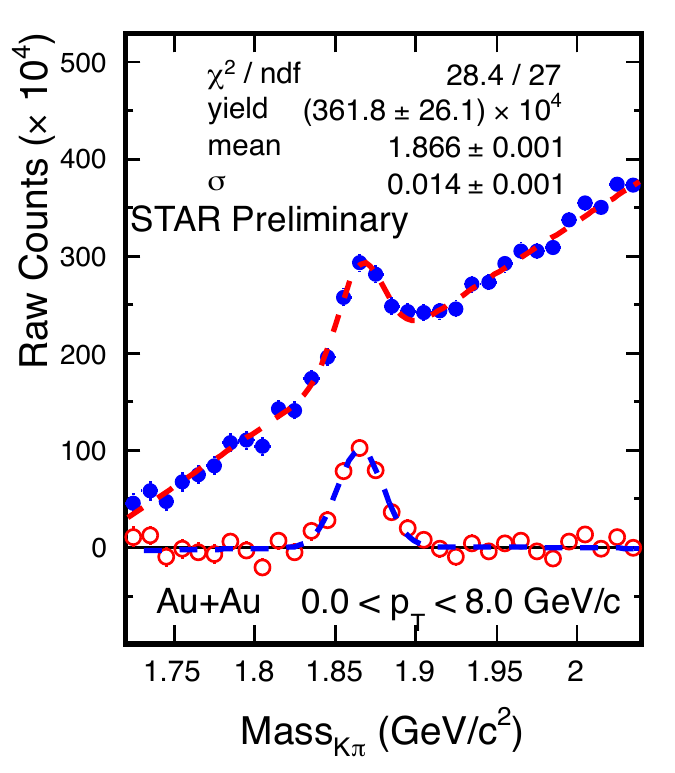}
\includegraphics[width=0.43\textwidth]{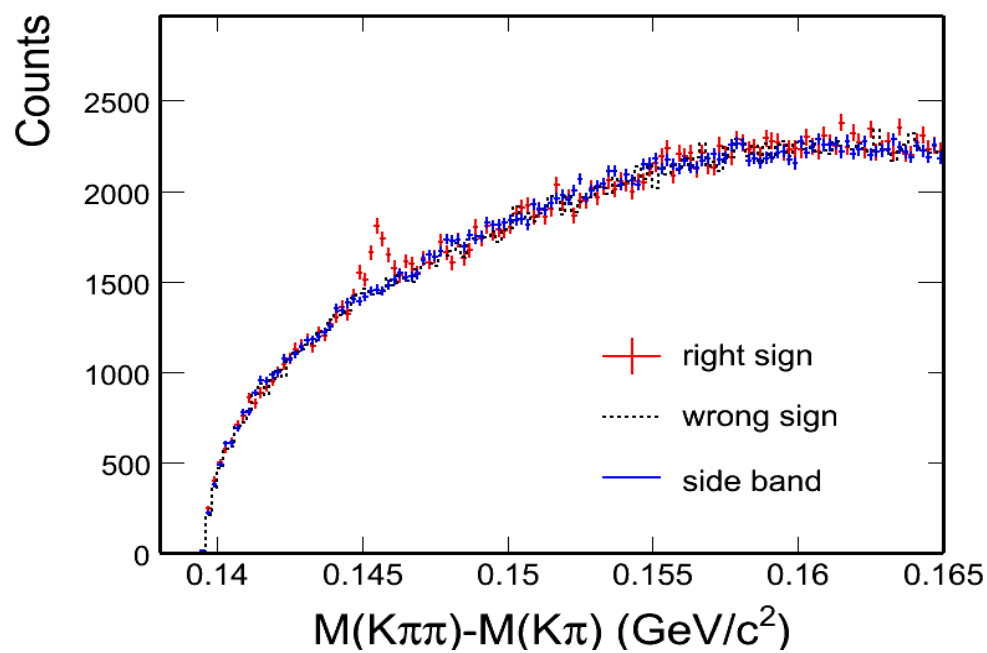}
\end{center}
\vspace{-7mm}
\caption{Left panel: $D^0$ signal in p+p 200 GeV  
collisions after same-sign background subtraction \cite{OpenCharm2011Paper}. Middle panel: $D^0$ signal in Au+Au 200 GeV  
collisions after mixed-event background subtraction. Right panel: $D^*$ 
signal in $p+p$ 200 GeV collisions with combinatorial background reproduced by wrong-sign and side-band methods \cite{OpenCharm2011Paper}. } 
\label{yields}
\end{figure}

$Y(p_T,y)$ is obtained from fitting the reconstructed invariant mass spectrum (Fig. \ref{yields}) of open charm mesons 
through hadronic decays: 
\begin{itemize}
\item $D^0(\overline{D^0})\rightarrow K^\mp\pi^\pm$ (BR = 3.89\%)
\item $D^{*\pm}\rightarrow D^0(\overline{D^0})\pi^\pm$ (BR = 67.7\%) $\rightarrow K^-\pi^+\pi^\pm$ (total BR = 2.63\%) 
\end{itemize}

The identification of daughter particles is done in the STAR experiment \cite{NIM} at
mid-rapidity $|y|<1$ at $\sqrt{s_{NN}}=200$ and 500 GeV. The analysis presented herein is done using three datasets; the first one collected in year 2009 ($N_\mathrm{trig}\sim$ 105 million 200 GeV p+p collisions), the second one collected in 2010 and 2011 ($N_\mathrm{trig}\sim$ 800~million Au+Au 200 GeV collisions), and the third one in 2011 ($N_\mathrm{MB}\sim$ 50 million 500 GeV p+p collisions).



At present, STAR does not have the capability to reconstruct the secondary vertex of $D^0$ decay; one must
calculate the invariant mass of all $K\pi$ pairs coming from the vicinity of the primary vertex. This results in a large 
combinatorial background which was reconstructed via the mixed-event method (Au+Au dataset), same-charge-sign, and kaon momentum-rotation ($p+p$ dataset) and subtracted from invariant mass spectra of all particle pairs \cite{OpenCharmPrev}. To reconstruct
$D^*$, one may exploit the softness of $D^*\rightarrow D^0\pi$ decay; combine low momentum pions with $D^0$
candidates, i.e. pairs with $1.82<M(K\pi)<1.9$ GeV/c$^2$, and plot difference $M(K\pi\pi)-M(K\pi)$ whose resolution
is determined by mostly the soft pion high momentum resolution.  
The combinatorial background is
reconstructed by side-band (picking  $K\pi$ pair outside the $D^0$ mass region) and wrong-sign (picking
soft pion with opposite charge) methods. The dominant source of systematic uncertainties for both $D^0$ and $D^*$ analyses
is the difference between yields obtained from subtractions of combinatorial background from all particle combinations.         

\vspace{-6mm}
\section{Results}
\vspace{-2mm}

\subsection{D meson production in $p+p$ collisions}
\label{par:pp}

Yields $Y(p_T,y)$ are calculated in six $p_T$ bins (first two for $D^0$, the next four for $D^*$) in $p+p$ 200 GeV and five $p_T$ bins (first for $D^0$, the next four for $D^*$) in $p+p$ 500 GeV. The charm cross section at mid-rapidity $\mathrm{d}\sigma^{c\overline{c}}/\mathrm{d}y$ was
obtained from power-law function 
\footnote[1]{$\frac{\mathrm{d}^2\sigma^{c\bar{c}}}{2\pi p_T\mathrm{d}p_T\mathrm{d}y} = 4\frac{\mathrm{d}\sigma^{c\bar{c}}}{\mathrm{d}y}\frac{(n-1)(n-2)}{\langle p_{T} \rangle^{2}(n-3)^{2}}\left(1+\frac{2p_T}{\langle p_T \rangle(n-3)}\right)^{-n}$}
fit to 
$\mathrm{d}^2\sigma^{c\overline{c}}/(2\pi p_T\mathrm{d}p_T\mathrm{d}y) = \mathrm{Inv}Y\cdot\sigma_\mathrm{NSD},
$ where $\mathrm{Inv}Y$ is obtained from (\ref{eq:InvariantYield}). $\sigma^{\mathrm{NSD}}$ is the total Non-single Diffractive (NSD) cross section, which is measured at STAR to be $30.0\pm2.4$~mb at $\sqrt{s}=200$ GeV \cite{StarNsd}. In the case of $\sqrt{s}=500$ GeV, there's no STAR measurement yet; $\sigma^{\mathrm{NSD}}$ is extrapolated from 200 GeV measurement with the help of PYTHIA simulation to be 34 mb. 
\begin{figure}[!h]
\begin{center}
\vspace{-5mm}
\includegraphics[width=0.42\textwidth]{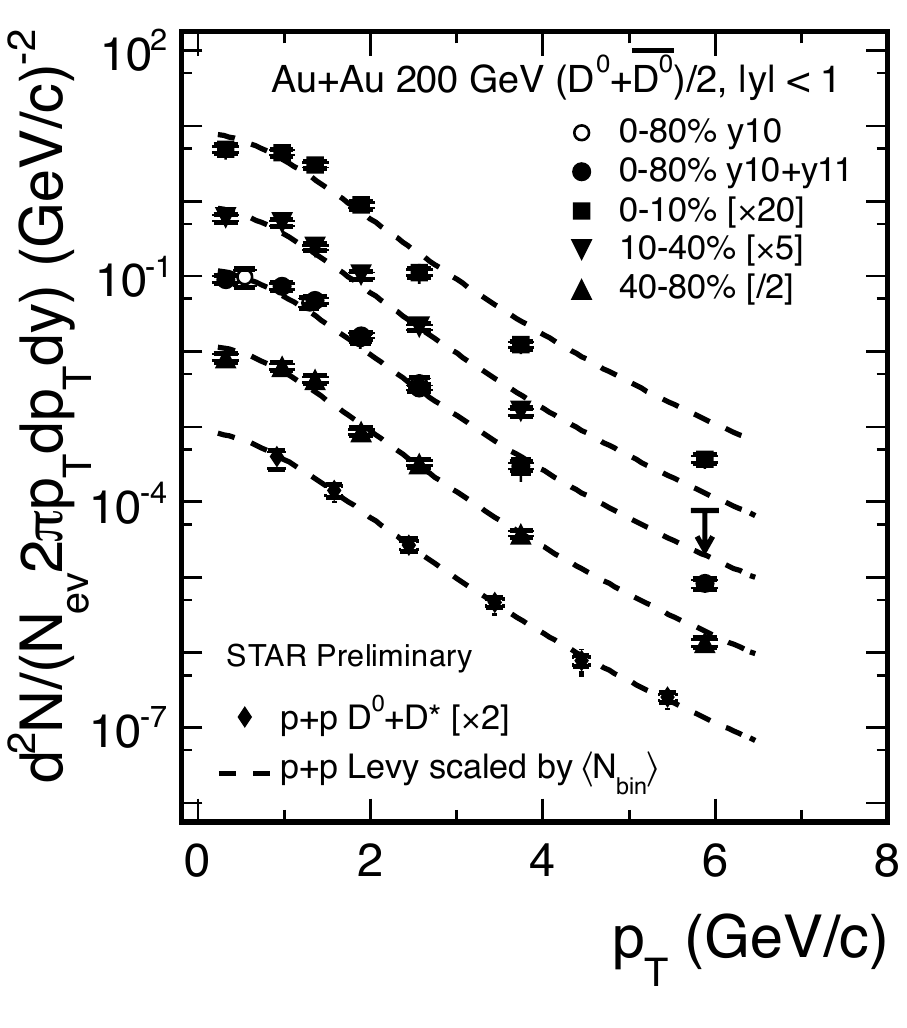}\qquad 
\includegraphics[width=0.478\textwidth]{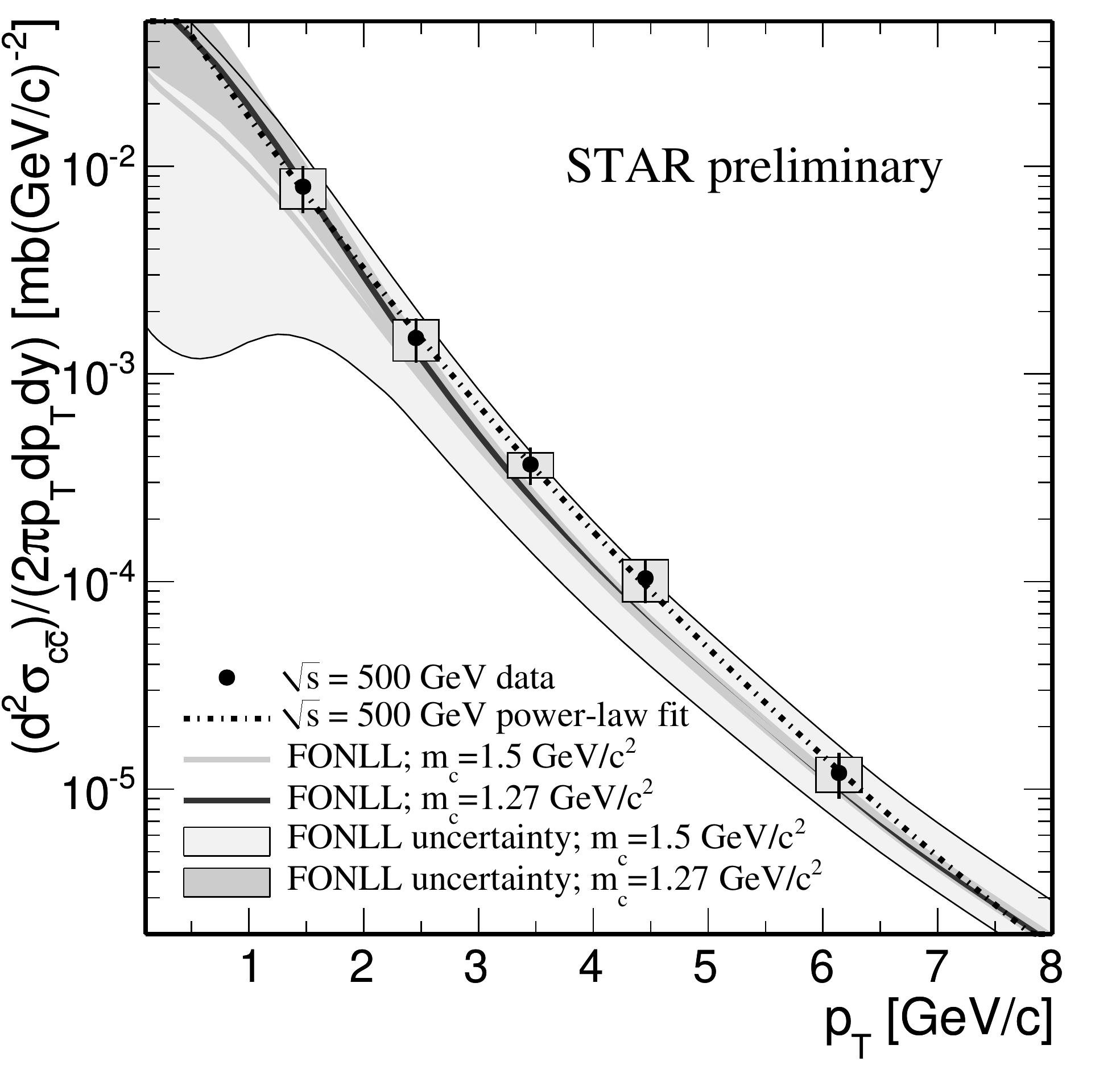} 
\end{center}
\vspace{-8mm}
\caption{Left Panel: $D^0$ Inv$Y$ spectra for various centralities, The last four $p_T$ bins in $p+p$ collisions are from $D^{+^*}$. Right Panel: Charm quark production invariant cross section as a function of D meson $p_T$ in 500 GeV p+p collisions with two FONLL predictions \cite{ramona} using normalization and factorization scale equal to charm quark mass $m_c$.}
\vspace{-5mm} 
\label{spectrum1}
\end{figure}
The charm production cross section at mid rapidity $\left . \frac{\mathrm{d}\sigma_{c\bar{c}}}{\mathrm{d}y}\right|_{y=0}$ is          $170\pm45(\mathrm{stat.})^{+37}_{-51}(\mathrm{sys.})\ \mu\mathrm{b}$ at $\sqrt{s} = 200$ GeV and is $ 217\pm86(\mathrm{stat.})\pm 73(\mathrm{sys.})\ \mu\mathrm{b}$ at $\sqrt{s} = 500$ GeV. FONLL predictions for $p_T$ spectra \cite{ramona} shown in Fig. \ref{spectrum1}. 


\subsection{$D^0$ production in Au+Au collisions}

\begin{figure}[!h]
\begin{center}
\vspace{-3mm}
\includegraphics[width=0.33\textwidth]{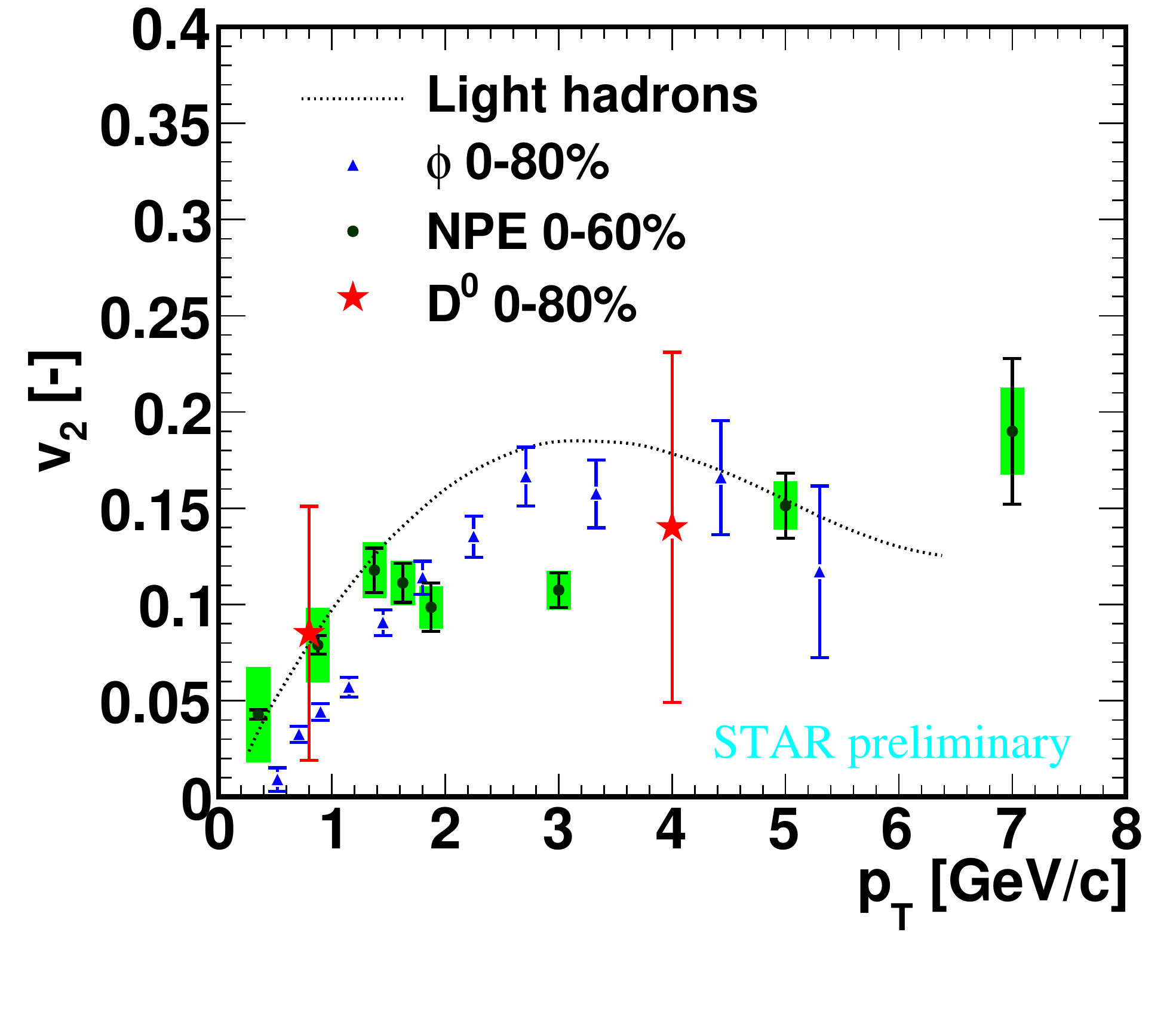} 
\includegraphics[width=0.352\textwidth]{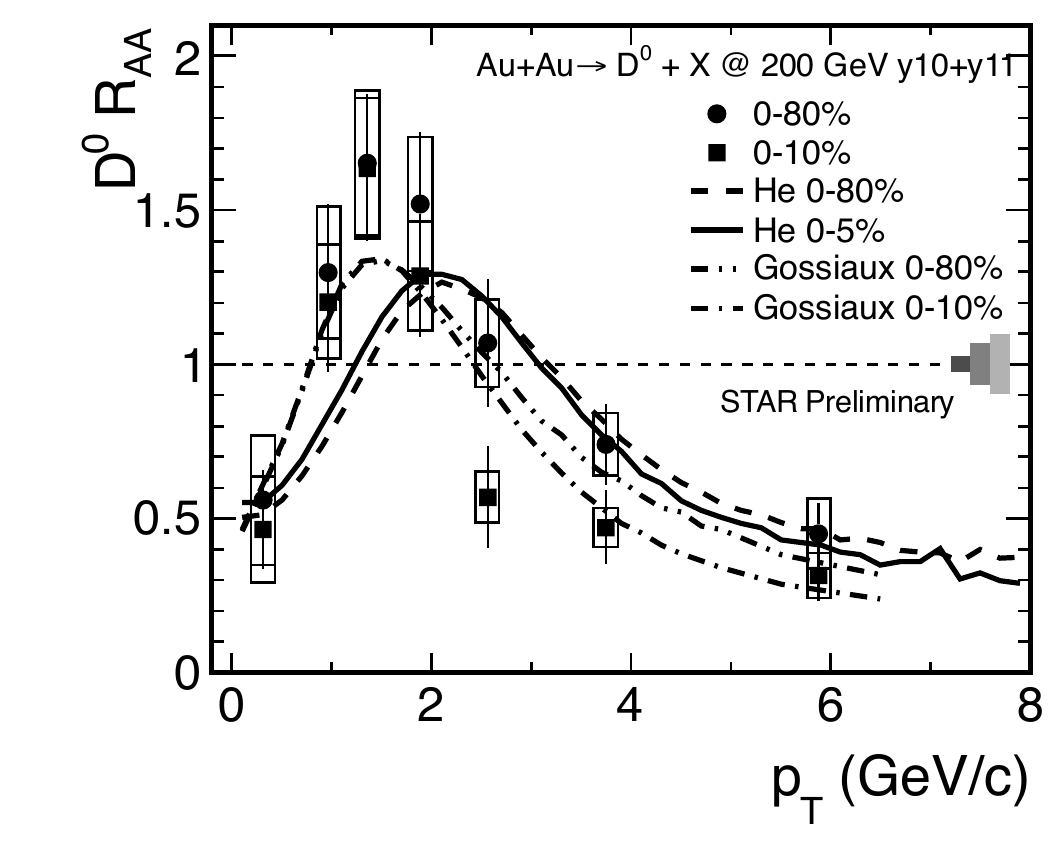} 
\includegraphics[width=0.30\textwidth]{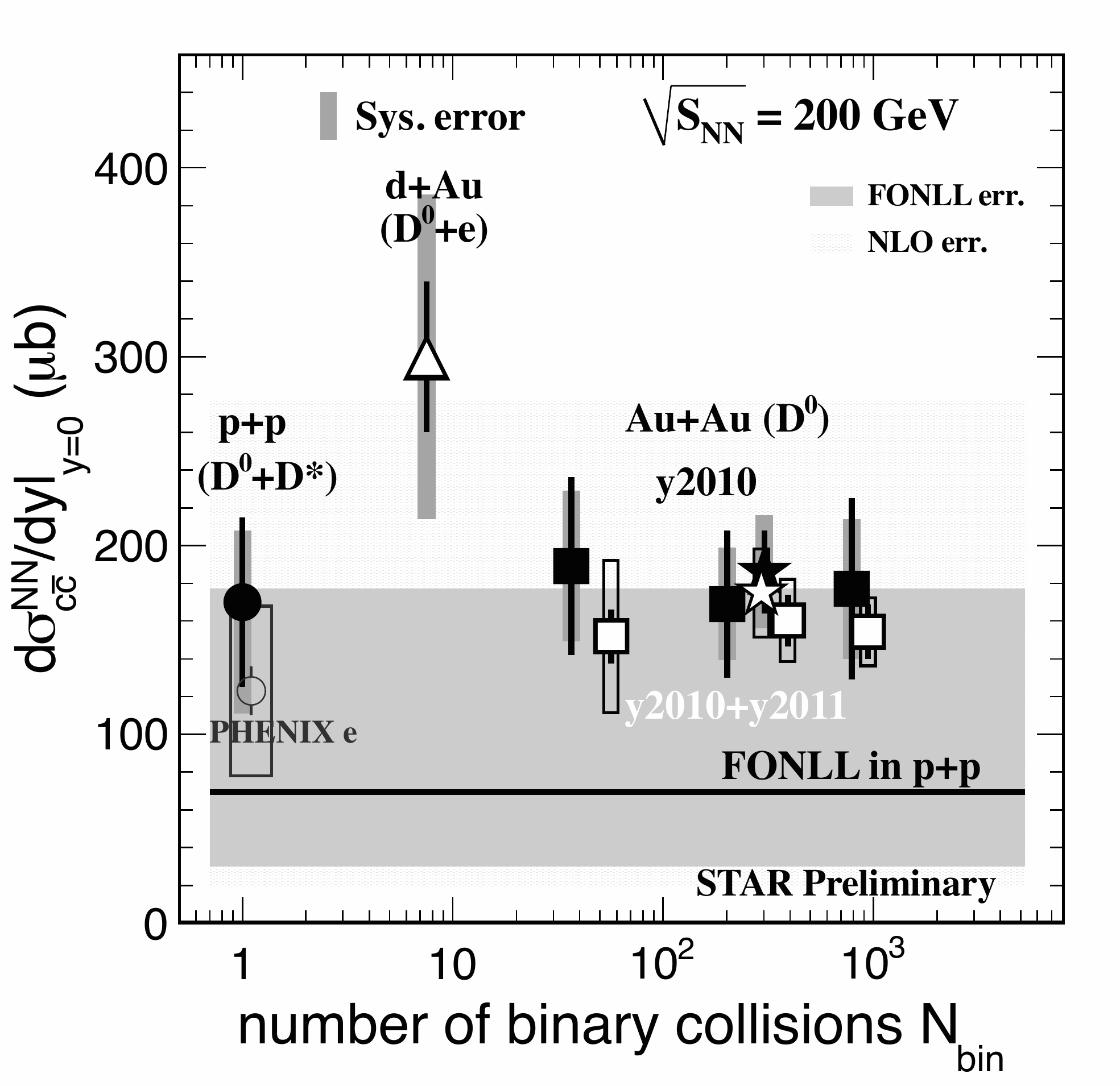} 
\end{center}
\vspace{-8mm}
\caption{Left Panel: Elliptic flow as a function of $p_T$ . Middle panel: $D^0$ nuclear modification factor $R_{\mathrm{AA}}$ as a function of $p_T$ for most central (blue) and minimum-bias (red) Au+Au collisions with theoretical predictions from two models \cite{He,Gossiaux}. Green rectangles around unity represent systematic uncertainties, from left to right, $N_\mathrm{bin}$ definition uncertainty for the most central (2.8\%), $N_\mathrm{bin}$ definition uncertainty for all Au+Au (7\%), and $p+p$ normalization error (8.1\%). Right panel: The charm production cross section per $N_\mathrm{bin}$ as a function of $N_\mathrm{bin}$.}
\label{spectrum2}
\end{figure}
\vspace{-3mm}

Yields $Y(p_T,y)$ were calculated in eight $p_T$ and three centrality bins. $\mathrm{d}\sigma_{c\overline{c}}^{\mathrm{NN}}/\mathrm{d}y$ was obtained from 
the integral of
\begin{equation}
\mathrm{d}^2\sigma_{c\overline{c}}^{\mathrm{NN}}/(2\pi p_T\mathrm{d}p_T\mathrm{d}y) = \mathrm{Inv}Y\cdot\sigma^\mathrm{inel}/N_\mathrm{bin}
\label{eq:AuAuXsection}
\end{equation}
over $p_T$ and is measured to be $\left . \frac{\mathrm{d}\sigma_{c\bar{c}}}{\mathrm{d}y}\right|_{y=0} = 175\pm13(\mathrm{stat.})\pm23(\mathrm{sys.})\ \mu\mathrm{b}$. 
$\mathrm{Inv}Y$ is obtained from (\ref{eq:InvariantYield}) and $\sigma^{\mathrm{inel}} = 42$ mb is the total inelastic cross section \cite{Honda}.  
To calculate the $D^0$ nuclear modification factor $R_\mathrm{AA}$ in various centrality bins, we scaled Levy function 
\footnote[2]{$\frac{1}{2\pi p_T}\frac{\mathrm{d}^2\sigma_{c\bar{c}}}{\mathrm{d}p_T\mathrm{d}y}=\frac{\mathrm{d}\sigma_{c\bar{c}}}{\mathrm{d}y}\frac{(n-1)(n-2)}{2\pi nC[nC+m_0(n-2)]}\left(1+\frac{\sqrt{p_T^2+m_0^2}-m_0}{nC}\right)^{-n}$}
fit to $p+p$ data by $N_\mathrm{bin}$, as shown in the left panel of Fig. \ref{spectrum1}, and follow the same process for the original power-law function as discussed in section \ref{par:pp}. Since enhanced statistics allow more $p_T$ bins in Au+Au collisions, we rely on the extrapolation from the two fits to estimate one source of systematic uncertainty. The $p+p$ baseline for $R_\mathrm{AA}$ calculation is the arithmetic average of the Levy and the power-law fit results. The measurement, shown in the middle panel of Fig. \ref{spectrum2}, reveals strong suppression in the most central collisions for $p_T > 2$ GeV/c consistent with the
prediction of the SUBATECH group (Gossiaux) model \cite{Gossiaux} and exhibits the maximum of the $R_\mathrm{AA}$ around $p_T\simeq 1.5$ GeV/c. This agreement with \cite{Gossiaux} might indicate that the maximum is 
induced by the transverse flow picked up from the expanding medium through coalescence with light-quarks.

d$\sigma^{\mathrm{NN}}_{c\bar{c}}/\mathrm{d}y|_{y=0}$ as a function of $N_{\mathrm{bin}}$ is shown in 
the right panel of Fig. \ref{spectrum2}. Within errors, the results are in agreement and follow the 
number-of-binary-collisions scaling, which indicates that charm quark is produced via 
initial hard scatterings at early stage of the collisions at RHIC. The FONLL (darker 
band) and NLO \cite{nlo} (lighter band) uncertainties are also shown here for comparison.

In the Left panel of Fig. \ref{spectrum2}, the measurement of $D^0$ elliptic flow $v_2$ is shown. Within large statistical
error bars, $D^0$ $v_2$ is consistent with the STAR Non-photonic electrons $v_2$ indicating non zero elliptic flow of $D^0$ mesons in Au+Au collisions at $\sqrt{s_\mathrm{NN}}$ = 200 GeV. 

\vspace{-3mm}
\section{CONCLUSIONS}
\vspace{-2mm}

New open charm hadrons $(D^0, D^{*+})$ measurements in $p+p$ and Au+Au minimum bias collisions at $\sqrt{s_{\mathrm{NN}}}=200$ GeV from STAR shows the $N_\mathrm{coll}$ scaling of the charm quark production cross section at mid rapidity. 
$\mathrm{d}\sigma^\mathrm{NN}_{c\overline{c}}/\mathrm{d}y|_{y=0}=170\pm45(\mathrm{stat.})^{+37}_{-51}(\mathrm{sys.})\ \mu\mathrm{b}$ in $p+p$, $175\pm13(\mathrm{stat.})\pm23(\mathrm{sys.})\ \mu\mathrm{b}$ in Au+Au collisions at 200 GeV  and $217\pm86(\mathrm{stat.})\pm 73(\mathrm{sys.})\ \mu\mathrm{b}$ in $p+p$ collisions at 500 GeV. 

The new $D^0$ nuclear modification factor $R_\mathrm{AA}$ measurement reveals strong suppression in the most central collisions for $p_T > 2$ GeV/c consistent with the
prediction of the model \cite{Gossiaux} and exhibits the maximum of the $R_\mathrm{AA}$ around $p_T\simeq 1.5$ GeV/c.     

In the near future the STAR Heavy Flavor Tracker \cite{hft} will provide the 
necessary resolution to reconstruct secondary vertices of charm mesons, which will increase 
the precision of charm measurements. 

\vspace{2mm}
{\small
\textbf{Acknowledments}

This work was supported by grant INGO LA09013 of the Ministry of Education, Youth and Sports
of the Czech Republic, and by the Grant Agency of the Czech Technical University in Prague, grant No.
SGS10/292/OHK4/3T/14.
}




\vspace{-4mm}

\bibliographystyle{elsarticle-num}
\bibliography{<your-bib-database>}



\end{document}